\documentclass[a4paper,twoside]{article}
\newcommand{\affil}[1]{$^{\rm #1}$}
\usepackage[authoryear]{natbib}
\bibpunct{(}{)}{;}{a}{}{,}
\usepackage{graphicx}
\date{} %Please leave the date blank
\newcommand{\radmsq}{\mbox{rad\,m$^{-2} $}}
\title{\large\bf\flushleft Detection Thresholds and Bias Correction in Polarized Intensity}
\author{\parbox{\textwidth}{\flushleft
\vspace{-0.5cm}
{\it Samuel J. George\affil{A}, Jeroen M. Stil\affil{A}, and Ben W. Keller\affil{A}}\\
\vspace{0.4cm}
{\small \affil{A}\,Institute for Space Imaging Science \& Department of Physics and Astronomy, The University of Calgary, 2500 University Drive NW, Calgary AB, T2N 1N4, Canada  \,Email: samuel@ras.ucalgary.ca}\\
}}
\begin{document}
\twocolumn[
\begin{changemargin}{.8cm}{.5cm}
\begin{minipage}{.9\textwidth}
\vspace{-1cm}
\maketitle
%
%%%%%%%%%%%%%     ABSTRACT    %%%%%%%%%%%%%
%Abstract of no more than 200 words here.  \small{\bf Abstract:}
Detection thresholds in polarized intensity and polarization bias
correction are investigated for surveys where the polarization
information is obtained from RM synthesis. 
Considering unresolved sources with a single rotation
measure, a detection threshold of $8 \sigma_{QU}$ applied to the Faraday
spectrum will retrieve the RM with a false detection rate less than
$10^{-4}$, but polarized intensity is more strongly biased than Ricean
statistics suggest. For a detection threshold of $5 \sigma_{QU}$, the
false detection rate increases to $\sim 4\%$, depending also on
$\lambda^2$ coverage and the extent of the Faraday spectrum.  Non-Gaussian
noise in Stokes $Q$ and $U$ due to imperfect imaging and calibration can
be represented by a distribution that is the sum of a Gaussian and an
exponential. The non-Gaussian wings of the noise distribution
increase the false detection rate in polarized intensity by orders of magnitude.
Monte-Carlo simulations assuming non-Gaussian noise in $Q$ and $U$, 
give false detection rates at $8 \sigma_{QU}$ similar to Ricean false detection 
rates at $4.9 \sigma_{QU}$.

%%%%%%%%%%%%%     KEYWORDS    %%%%%%%%%%%%%
\medskip{\bf Keywords:} polarization --- methods: statistical --- methods: data analysis
% Please write all keywords in lower case. PASA uses the
% standard list of subject headings adopted by The Astrophysical Journal
% and available from http://www.journals.uchicago.edu/ApJ/keywords_text.html.
% Keywords are separated by em-dashes, i.e. ---
%%%%%%%%DO NOT EDIT%%%%%%%%%%%%
\medskip
\medskip
\end{minipage}
\end{changemargin}
]
\small
%%%%%%%%EDIT FROM HERE%%%%%%%%%%%%

\section{Introduction}

Linear polarization of radio sources contains information on magnetic
fields in these sources, and Faraday rotation of the plane of
polarization provides information on the direction and magnitude of
the magnetic field along the line of sight. As such, observations of
linear polarization of radio sources provide the most widely
applicable probe of cosmic magnetic fields on scales from galaxies to
clusters of galaxies. Finding polarized sources in survey images and
fitting their parameters forms the basis of this analysis.

Sources with detectable polarized emission are readily identified in
images of total intensity. As a first approximation, source finding in
polarization can be reduced to applying a suitable detection threshold
to the polarized intensity at the location of every radio source
identified in total intensity. In practice, source finding in
polarization is more complicated because of two reasons.  The first is
related to resolved sources, and the second is related to Faraday
rotation of the polarized emission.

\begin{figure}[h]
\begin{center}
\includegraphics[scale=0.42, angle=0]{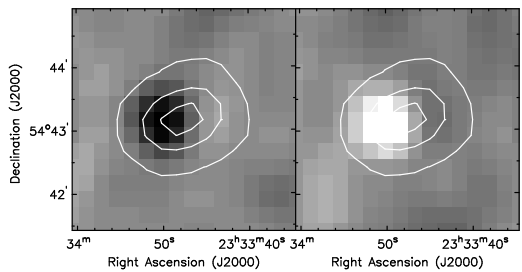}
\caption{A radio source that is resolved in total intensity, 
with one-sided polarization. Grey scales show Stokes $Q$ and 
$U$ from the NVSS with contours of total intensity at $3$, $30$, 
and $60$~mJy. The grey scales range from $-3$~mJy (black) to 
$+3$~mJy (white).}\label{NVSS_resolved-fig}
\end{center}
\end{figure}

Figure~\ref{NVSS_resolved-fig} shows a radio source from the 
National Radio Astronomy Observatory (NRAO) Very Large Array (VLA)
Sky Survey \cite[NVSS;][]{condon1998} that is slightly resolved in
total intensity (white contours), with major axis $68$'' in position
angle $-72^\circ$. The polarized emission is shown in grey scales as
Stokes $Q$ and $U$ images. The source has a component that is unresolved
at the $45$'' resolution of the NVSS. The unresolved component has a
peak polarized intensity of $6$~mJy ($\sigma_{QU}$ = $0.35$~mJy),
approximately twice the catalogued value derived from the polarized
intensity at the fitted position of the source in total intensity. It
is not clear which fraction of radio sources has different morphology
in polarized intensity than in total intensity.  Approximately $8$\% of
NVSS sources brighter than $10$~mJy outside the Galactic plane ($|b| >
30^\circ$) has a fitted (i.e. before deconvolution) major axis size
more than $1.5$ times the $45''$ (FWHM) size of the synthesized beam,
and $2\%$ has a fitted major axis more than twice the beam size.

Resolved polarized sources have been treated in different ways in the
literature.  The NVSS catalog derived polarized intensity at the
location of the fitted position in total intensity, and the listed
polarized flux density (peak times Stokes $I$ solid angle) implicitly
assumes a constant polarization angle over the source. 
\citet{taylor2007} and \citet{grant2010} fitted 2-dimensional
Gaussians to sources in polarized intensity. \citet{Subrahmanyan2010}
integrated Stokes $Q$ and $U$ over the solid angle of the source defined
in a low-resolution total intensity image, and catalogued each
polarized source as if it were unresolved.

The second complication for source finding in linear polarization is 
presented by Faraday rotation, even if the source is unresolved.  
Faraday rotation rotates the polarization angle $\Psi$ by an amount 
proportional to $\lambda^2$ for a simple Faraday thin source, and in a 
more complicated manner e.g. if synchrotron emission and Faraday rotation 
both occur in the same volume. Differential Faraday rotation over the
observed frequency range leads to depolarization, thus introducing a
selection effect against sources with strong Faraday rotation
\citep[][for the NVSS]{stil2007}. Multifrequency observations allow
Rotation Measure (RM) Synthesis \cite[][]{burn1966, brentjens2005} 
to solve for the unknown Faraday depth, polarized intensity, and
polarization angle simultaneously. The source is typically identified
by the maximum value in the Faraday spectrum, e.g. through the RM clean
algorithm \citep{heald2009}. 

Following \cite{brentjens2005} the Faraday depth $\phi$ is defined as

\begin{equation}
\phi\left(\vec{r}\right) = 0.81 \int_{0}^{x} \vec{B_{||}} n_{e} \cdot \mathrm{d}\vec{r}\ \mbox{rad}\ \mbox{m}^{-2},
\end{equation}

where $\vec{B_{||}}$ is the line of sight magnetic field component, $n_e$ is the 
thermal electron density, $\mathrm{d}\vec{r}$ is an infinitesimal path length, 
with the integral taken from the observer to the point $x$.

The complex polarized intensity  $\mathcal{P}\left(\lambda^{2}\right) = Q + iU$ is the Fourier transform of the
Faraday dispersion function $F(\phi)$,

\begin{equation}
P\left(\lambda^{2}\right) = \int_{-\infty}^{\infty} F(\phi) e^{2 i  \phi \lambda^{2}} d\phi .
\end{equation}

The Faraday rotation measure $RM$ is defined as the slope of 
a polarization angle $\Psi$ versus $\lambda^{2}$ plot:

\begin{equation}
RM\left(\lambda\right) = \frac{d \Psi}{d(\lambda^{2})} 
\end{equation}

where
\begin{equation}
\Psi = \frac{1}{2} \tan^{-1}\frac{U}{Q}.
\end{equation}

Once a polarized source has been detected, the observed polarized
intensity $p = \sqrt{Q^{2} + U^{2}}$ must be corrected for polarization
bias. Since $p$ is a positive-definite quantity, the noise in Stokes $Q$
and $U$ results in a positive value for $p$ even if no signal is
present. The statistics of $p$ with a signal $p_0$ and noise
$\sigma_{QU}$ is given by the Rice distribution
\citep{rice1945}. Estimators of the true polarized flux density $p_0$
from the observed polarized flux density $p$ based on this
distribution have been discussed by a number of authors
\citep{simmons1985,vaillancourt2006}. The Rice distribution assumes
Gaussian noise in Stokes $Q$ and $U$. Real surveys of the sky may have
non-Gaussian tails to the noise distribution resulting from imperfect
imaging and calibration. In this paper we investigate the detection
statistics in polarized intensity and polarization bias correction for
polarized intensity determined from RM synthesis and
in the presence of non-Gaussian noise in Stokes $Q$ and $U$.

Future radio polarization surveys such as the Galactic Arecibo L-band 
Feed Array Continuum Transit Survey \cite[GALFACTS;][]{taylor2010} 
and the Polarization Sky Survey of the Universe's Magnetism \\
\cite[POSSUM;][]{gaensler2010} are wide band, multifrequency surveys 
that require revision of detection threshold and polarization bias correction.

\section{RM Synthesis simulations}

\begin{table}[h]
\begin{center}
\caption{False detection rates in simulated Faraday spectra. 
(1) Ratio of true polarized signal to the
noise in Stokes $Q$ and $U$. (2) Percentage of simulated spectra with
$|~\phi-\phi_{0}~| > 50\ \radmsq$. No false detections were found in the $3.33
\times 10^4$ $8\sigma_{QU}$ simulated spectra. These percentages
depend in part on the Faraday depth range considered, and the width of the 
RM spread function. }\label{FDR_RM_table} 
\begin{tabular}{cc}
\hline $p_0/\sigma_{QU}$ & \% False Detection \\
     (1)            &   (2)\ \   \\
\hline 
2.0 & 73.9 \\
3.0 & 43.9 \\
4.0 & 16.7 \\
5.0 & \ 3.6 \\
6.0 & 0.43 \\
7.0 & 0.033 \\
8.0 & \ldots\\
\hline
\end{tabular}
\medskip\\
%%$^a$Table footnotes go here.\\
\end{center}
\end{table}

RM synthesis was performed on simulated data for sources with signal
to noise ratio $p_0/\sigma_{QU}$ ranging from $0$ to $15$. Each
realization contained a source with the prescribed polarized signal
and random polarization angle at the reference frequency, $1400$~MHz. 
Stokes $Q$ and $U$ values were calculated assuming a Faraday depth 
of $150$ \radmsq~  for $N_{\rm chan} = 1024$ frequency channels between
$1000$~MHz and $1400$~MHz. Gaussian noise with standard deviation $\sigma
= \sqrt{N_{\rm chan}}$ was added to each channel, resulting in a
standard deviation $\sigma_{QU} = 1$ for the noise in Stokes $Q$ and $U$
after averaging over all channels. RM synthesis was then performed on
the synthetic spectrum of complex polarization. The effect of 
spectral index $\alpha$ ($S_\nu \sim \nu^\alpha$),
assuming that the percentage polarization is constant across the band,
was investigated with separate simulations for $\alpha = 0$ and
$\alpha = -0.75$. For each combination of $p_0/\sigma_{QU}$ and
$\alpha$, $33\,\,300$ simulated Faraday spectra were analyzed.

\begin{figure}[h]
\begin{center}
\includegraphics[scale=0.42, angle=0]{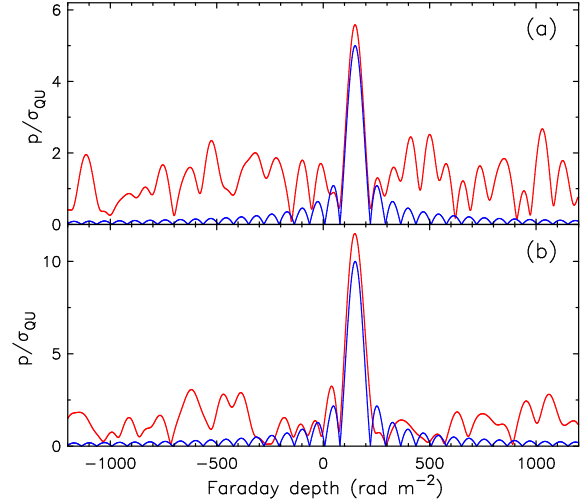}
\caption{Faraday spectra of simulated sources with signal to noise ratio
 $p_0/\sigma_{QU} = 5$ (a) and $10$ (b), both with Faraday depth of $150$
 \radmsq. The red curves represent the spectrum with noise, while the
 blue curves show the corresponding spectrum without
 noise.}\label{RM_spec-fig}
\end{center}
\end{figure}

Figure~\ref{RM_spec-fig} shows two simulated Faraday spectra with and
without noise. The noiseless spectra show the RM spread function with
its side lobes. The near side lobes of the RM spread function raise
the probability of a false peak at the wrong Faraday depth, resulting in stronger
wings in the error function of the Faraday depth when the signal to noise ratio is
low. As the Faraday depth of the source is not known a priori, the location of
the peak in the Faraday spectrum is catalogued as the Faraday depth of the source, and
the amplitude of the peak as the observed polarized intensity
$p_{\rm max}$.  We also extract the polarized intensity $p$ at the input
Faraday depth, because of the expectation that the Rice distribution with noise
$\sigma_{QU}=1$ applies to $p$, not necessarily to $p_{\rm max}$.

\begin{table*}[h]
\begin{center}
\caption{Expectation values of polarized intensity and bias correction.
(1) Spectral index ($S_\nu \sim \nu^\alpha$). (2) Input
polarized intensity at centre of the band in units of $\sigma_{QU}$,
and constant percentage polarization across the frequency band. (3)
Effective polarized intensity defined in Equation 2. (4) Polarized intensity 
in the Faraday profile at the input Faraday depth ($150\ \radmsq$). This
quantity is not known for real sources. (5) Maximum polarized intensity 
taken over all Faraday depth values. (6) Estimator of $p_{0,eff}$ taking polarized intensity from column 4, 
according to $\hat{p}_{\phi_0} = \sqrt{p_{\phi_0}^2 - \sigma_{QU}^2}$. This
quantity is not known for real sources. (7) Estimator of $p_{0,eff}$ taking polarized intensity from column 5,
according to $\hat{p}_0 = \sqrt{p_{\rm max}^2 - 2.3 \sigma_{QU}^2}$.}\label{expect_table} 
\begin{tabular}{rrrrrrr}
\hline $\alpha$ & $p_0/\sigma_{QU}$ & $p_{0,eff}/\sigma_{QU}$ & $p_{\phi_0}/\sigma_{QU}$ &  $p_{\rm max}/\sigma_{QU}$ & $\hat{p}_{\phi_0}/\sigma_{QU}$ & $\hat{p}_0/\sigma_{QU}$   \\
   (1)   &  (2)      &      (3)    &   (4)    &     (5)  &  (6)    &  (7)     \\
\hline
 0.00  &    4.000  &     4.000  &   4.139  &    4.384 &  4.020  & 4.114     \\
 0.00  &    5.000  &     5.000  &   5.106  &    5.237 &  5.009  & 5.012     \\ 
 0.00  &    6.000  &     6.000  &   6.092  &    6.188 &  6.011  & 5.999     \\   
 0.00  &    7.000  &     7.000  &   7.076  &    7.156 &  7.006  & 6.994     \\  
 0.00  &    8.000  &     8.000  &   8.072  &    8.142 &  8.011  & 8.000     \\  
 0.00  &    9.000  &     9.000  &   9.070  &    9.130 &  9.015  & 9.004     \\  
 0.00  &   10.000  &    10.000  &  10.053  &   10.107 & 10.003  & 9.993      \\ 
 0.00  &   15.000  &    15.000  &  15.027  &   15.063 & 14.994  & 14.986      \\ 
\hline
 $-$0.75 &    4.000 &    4.110 &   4.246  &   4.467  &  4.130  & 4.202      \\ 
 $-$0.75 &    5.000 &    5.138 &   5.244  &   5.365  &  5.150  & 5.147      \\ 
 $-$0.75 &    6.000 &    6.165 &   6.255  &   6.345  &  6.175  & 6.161      \\ 
 $-$0.75 &    7.000 &    7.193 &   7.272  &   7.347  &  7.204  & 7.188      \\ 
 $-$0.75 &    8.000 &    8.221 &   8.288  &   8.353  &  8.227  & 8.215      \\ 
 $-$0.75 &    9.000 &    9.248 &   9.305  &   9.362  &  9.251  & 9.238      \\ 
 $-$0.75 &   10.000 &   10.276 &  10.332  &  10.383  & 10.284  & 10.272      \\ 
 $-$0.75 &   15.000 &   15.413 &  15.444  &  15.478  & 15.412  & 15.403      \\ 
\hline
\end{tabular}
\medskip\\
%%$^a$Table footnotes go here.\\
\end{center}
\end{table*}

\begin{figure}[h]
\begin{center}
\includegraphics[scale=0.42, angle=0]{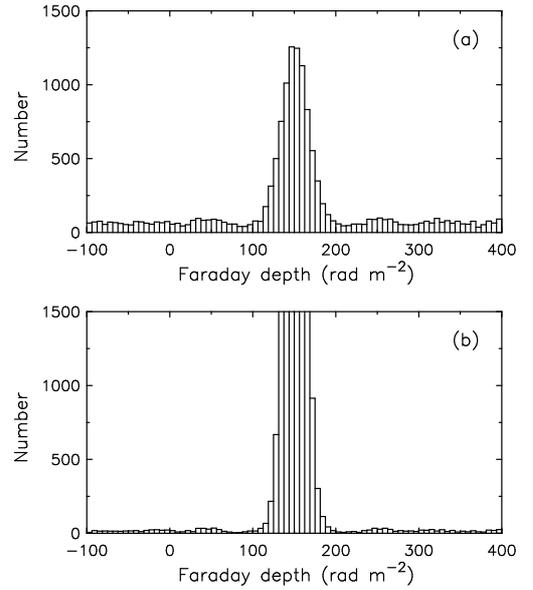}
\caption{Distribution of Faraday depths, with input $\phi_{0}$ $= 150\ \radmsq$, 
 derived from simulations with (a)  $p/\sigma_{QU} = 2$ 
 and (b) $p/\sigma_{QU} = 4$.}\label{RM_err-fig}
\end{center}
\end{figure}

Figure~\ref{RM_err-fig} shows the distribution of Faraday depths derived from 
the simulations for $p/\sigma_{QU} = 2$ and $4$ and $\alpha=0$.
While each Faraday profile contains a source with the input $\phi_{0}$ $=
150\ \radmsq$, the uniform distribution of Faraday depths at low signal to noise
ratios represents spurious peaks in excess of the actual source.  The
effect of these false detections is twofold: expectation value of 
the polarized intensity approximates a nearly constant value for 
$p_0/\sigma_{QU} < 3$, and the error distribution in Faraday depth becomes 
significantly non-Gaussian.  Table \ref{FDR_RM_table} lists false detection 
rates as a function of signal to noise ratio. In the range 
$3 \le p_0/\sigma_{QU} \le 8$ the false detection rate drops
to below $10^{-4}$, and the error distribution of Faraday depth becomes
approximately Gaussian.

\begin{figure}[h]
\begin{center}
\includegraphics[scale=0.42, angle=0]{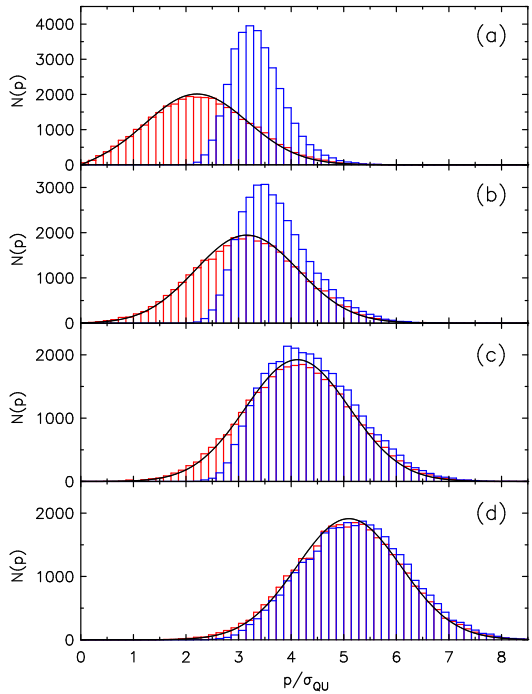}
\caption{Distribution of polarized intensity derived from simulated Faraday spectra. 
 Signal to noise $p_{0}/\sigma_{QU}=2, 3, 4, 5$ for
 panels (a)-(d) respectively. The red histograms represent values of
 $p$ at the actual Faraday depth of the source, and the black curves show the
 Rice distribution for the assumed value of $p_{0}/\sigma_{QU}$. The
 blue histograms represent polarized intensity at the peak of the Faraday
 spectrum.}\label{RM_sim_p-fig}
\end{center}
\end{figure}

Figure~\ref{RM_sim_p-fig} shows the distributions of polarized
intensity derived from the simulations, along with curves representing
the Rice distribution for $p_0/\sigma_{QU} =$ $2$, $3$, $4$, and $5$.  
The red histogram shows the polarized intensity $p_{\phi_{0}}$ at the actual Faraday depth of 
the source, demonstrating the Ricean statistics for this quantity. In
reality, we do not know the actual Faraday depth of the source, but solve for
this by finding $p_{\rm max}$, defined as the peak of the Faraday 
spectrum. The distributions of $p_{\rm max}$ are shown by the blue
histograms in Figure~\ref{RM_sim_p-fig}. At any signal to noise ratio,
the distribution of $p_{\rm max}$ is shifted to higher $p$ with
respect to the Rice distribution, adding to the polarization
bias. This additional bias is closely related to the {\it fitting
 bias} for fitting the flux density of a source discussed by
\citet{condon1998}. The magnitude of this bias is similar to the
well-known polarization bias.

For $p_0/\sigma_{QU}<5$, the distribution of $p_{\rm max}$ is skewed
with respect to the Rice distribution. The distribution of
$p_{\rm max}$ can be approximated by the distribution of the maximum of
$N-1$ independent draws from the Rice distribution with no signal, and
$1$ draw from the Rice distribution with the seeded signal $p_{0}$, where
$N$ is the ratio of the range of the Faraday spectrum to the FWHM width of
the RM spread function. The sidelobes of the RM spread function
increase the false detection rate, so this explanation can only be an
approximation. The difference between the distribution of $p_{\rm max}$
and the Rice distribution at low signal to noise depends on the range
of the Faraday spectrum.  At high signal to noise ratios, the maximum is
always associated with the source, and the fitting bias is independent
of the range of the RM spectrum. 

Table~\ref{expect_table} lists expectation values of polarized
intensity for flat spectrum sources and steep
spectrum sources at a range of signal to noise ratios in polarized
intensity. Column 6 lists the estimated true polarized intensity 
using $\hat{p}^{\phantom{1}}_{\phi_{0}} = \sqrt{p_{\phi_{0}}^{2} - \sigma_{QU}^{2}}$.
Though this is not measurable in real data, the correspondence between 
column 6 and column 2 reflects the Ricean statistics of $p_{\phi_{0}}$ 
illustrated in Figure~\ref{RM_sim_p-fig}. The bias in $p_{\rm max}$ measured from real data 
is approximately twice as large as in  $p_{\phi_{0}}$.
The polarization bias correction can be adjusted to
correct for the additional bias associated with the uncertainty in
RM. The estimator
\begin{equation}
\hat{p}_0 = \sqrt{p^2 - 2.3 \sigma_{QU}^2}
\end{equation}
for $p/\sigma_{QU} > 4$ provides accurate estimates of $p_0$.

This work suggests that a detection threshold of $p_0/\sigma_{QU} > 8$
should be applied for the derivation of Faraday depth, in order to obtain a
well-behaved error function of Faraday depth. Polarized intensity can be
estimated for sources with $p_0/\sigma_{QU} > 4$, depending on the
desired level of acceptable false detections. 

In the case where $\alpha \ne 0$, \citet{brentjens2005} recommend dividing
by the total brightness as a function of frequency. This may work well for
bright sources, but not for faint sources, or diffuse polarized emission.
In our experiments, we find that the Faraday depth derived for sources 
with $\alpha = -0.75$ was not significantly different from sources with 
$\alpha = 0$, but the polarized intensity after polarization bias correction 
is given by $p_{eff}$ defined as
\begin{equation}
p_{eff} = {{\int p d\lambda^2} \over {\int d\lambda^2}},
\end{equation}
where the integral is evaluated over the wavelength range of the data.
The penalty of not dividing by total intensity creates a spectral-index dependent
bias in polarization that is larger than the effects discussed previously.

\begin{figure}[h]
\begin{center}
\includegraphics[scale=0.42, angle=0]{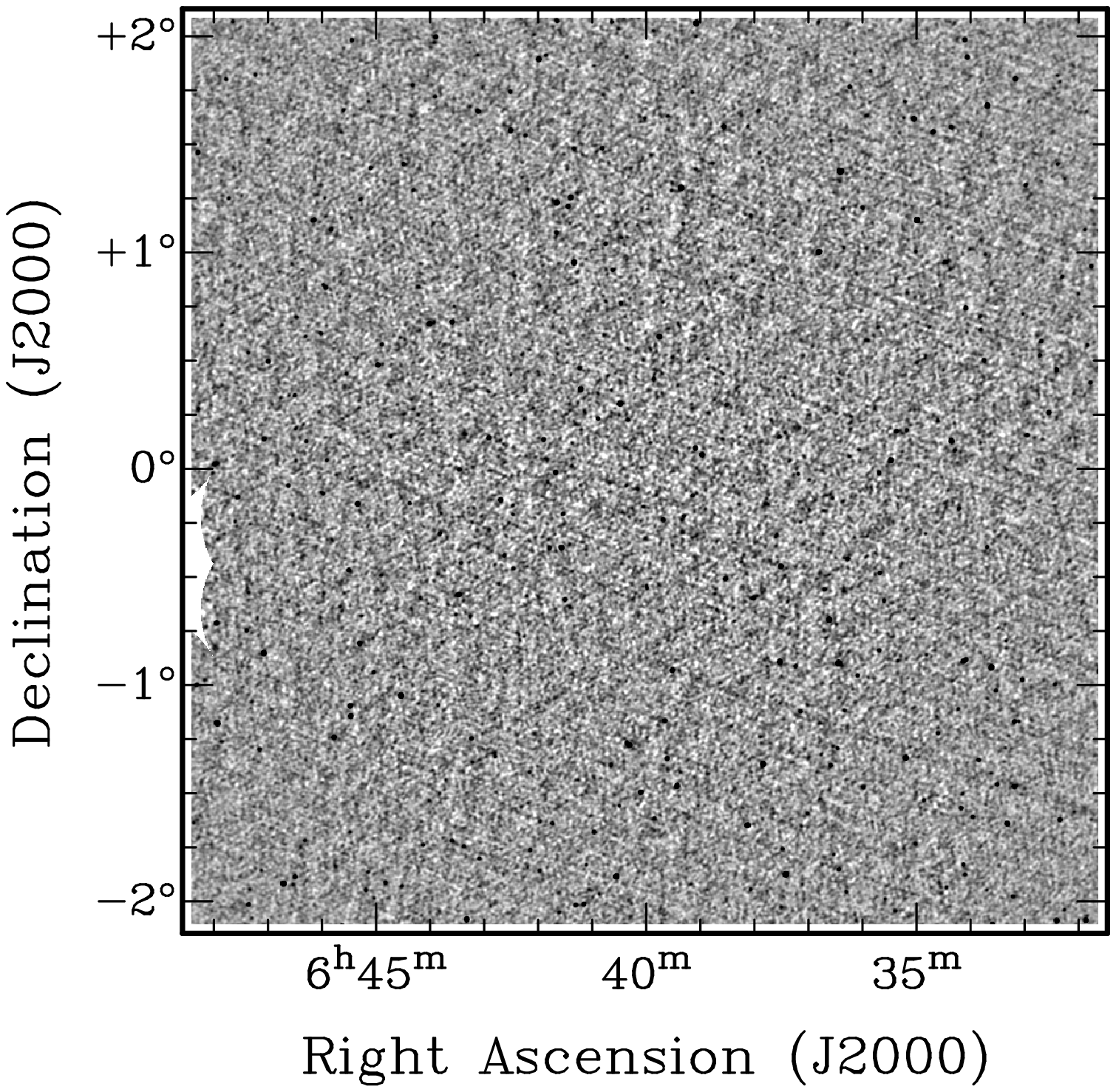}
\includegraphics[scale=0.42, angle=0]{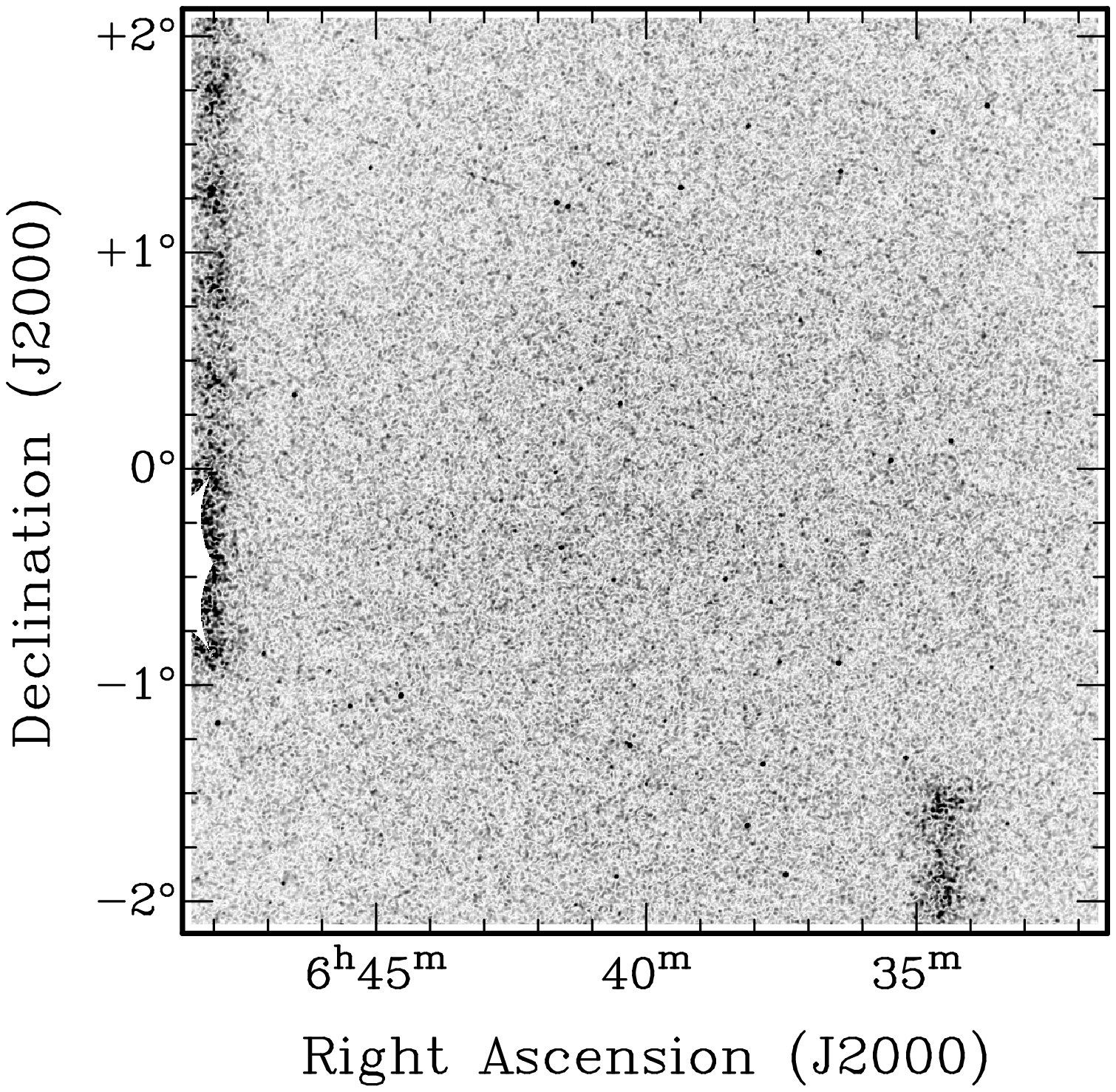}
\caption{Example of one of the $4^{\circ} \times 4^{\circ}$ fields in (top) 
total and (bottom) polarized intensity, illustrating the effects of 
missing fields on the noise.}\label{simulation-fig}
\end{center}
\end{figure}

\begin{figure}[h]
\begin{center}
\includegraphics[scale=0.40, angle=0]{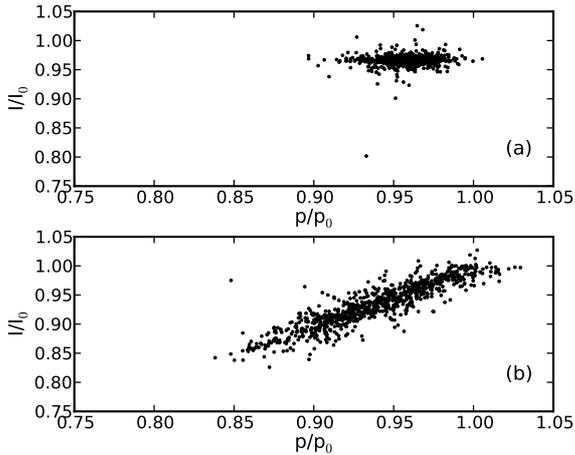}
\caption{Comparison of the ratio of peak values ($I$ and $p$) retrieved from the image 
and the input values ($I_{0}$ and $p_{0}$) for sources with $p > 20$~mJy.
(a) shows peak fluxes determined from fitting with SAD, 
(b) shows the nearest pixel value.}\label{pcompare-fig}
\end{center}
\end{figure}

\section{Simulated Sky Survey}

Sky simulations with sensitivity and angular resolution similar to the NVSS 
over an area of $2.33$~sr covering $478$ $4^{\circ} \times 4^{\circ}$ 
fields were constructed to test source finding and stacking algorithms. Figure \ref{simulation-fig} 
shows a simulated image in total intensity and polarized intensity. The images were 
constructed with source density and noise level similar to the NVSS polarization images. 
The images were built up by seeding sources at random positions in the image plane. 
Each source consists of a VLA snapshot antenna pattern scaled to the assumed clean 
limit plus a two dimensional Gaussian representing the restored clean components. 
The sidelobes of the antenna patterns from different sources add up incoherently 
simulating incomplete cleaning. The Gaussian noise divided by the sensitivity 
pattern of the NVSS mosaics was added to the Stokes $I$, $Q$ and $U$ images. 
The noise level in Stokes $I$ is $2$~mJy, whilst $Q$ and $U$ is $0.4$~mJy. 
Images with missing fields were included to simulate survey edges. 
The distribution of polarized fraction of sources followed the distribution 
derived by \cite{beck2004} for NVSS source brighter than $80$~mJy. 
The images only contain unresolved sources and the resolution of the 
images is $45$'' with a pixel size of $15$''.

The images were searched for sources in total intensity and
polarized intensity with the AIPS (Astronomical Image Processing System) source 
finder SAD (Search and Destroy), with a detection threshold of $5$~mJy. The recovered 
sources were matched with the input source catalogue. Only sources that matched 
within $60''$ were considered for further analysis. For sources with $I > 50$~mJy 
the standard deviations in right ascension and declination were $1.47$'' 
and $1.46$'' respectively. 

The peak flux in both total and polarized intensity is found in two ways. By using 
SAD to find and fit the sources and by extracting the nearest pixel value for $I$ and $p$. 
The nearest pixel values are considered because RM synthesis is done per pixel. 
Figure \ref{pcompare-fig} compares the fitted peak and the nearest pixel values with the 
input catalog for both total intensity and polarized intensity for sources with $p >20$~mJy. 
The fitted peak values underestimate the input values by a few percent, but the solid angle of the source 
from the fits is slightly overestimated so that the integrated flux density is retrieved 
from the catalogue. The nearest pixel intensities underestimate the true intensity by up 
to 15\%, approximately along the line of constant $p/I$. The error in the fitted position 
of the source is much smaller than a pixel, so the uncertainty in the position of the source 
in total intensity does not introduce a significant error in the estimation of $p$. 
RM synthesis on the brightest pixel would introduce a systematic error in polarized intensity 
comparable to that shown in Figure \ref{pcompare-fig}b.
Figure \ref{pcompare-fig} suggests that source finding in the image plane after RM synthesis 
is required to obtain polarized flux densities with an accuracy better than $\sim 10\%$. 

\section{Non-Gaussian noise in $Q$/$U$}

The statistics of polarized intensity are usually described by the Rice distribution 
that assumes Gaussian noise in $Q$ and $U$. Imperfect imaging and calibration result 
in images that do not achieve the theoretical noise levels. The actual noise distribution 
has strong wings above Gaussian noise. 

\begin{figure}[h]
\begin{center}
\includegraphics[scale=0.36, angle=0]{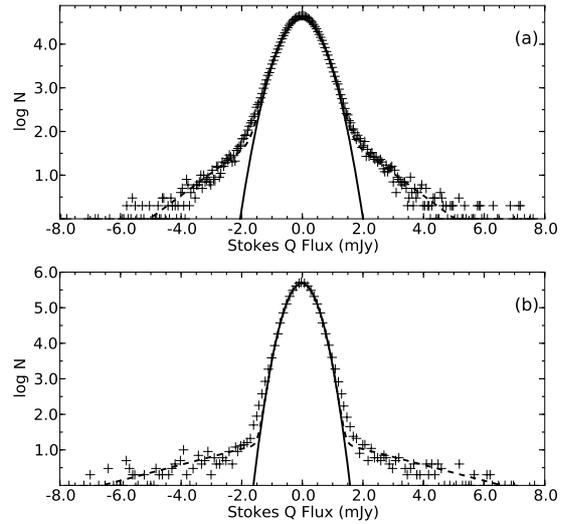}
\caption{Histograms of the noise for Stokes $Q$ images for (a) the simulated 
images (b) the NVSS. The fitted curves are (solid) a Gaussian fit and 
(dashed) the summation of a Gaussian and an exponential.}\label{noisehisto-fig}
\end{center}
\end{figure}

Figure \ref{noisehisto-fig} shows the noise distribution for both the simulated 
and the NVSS images. Sources identified in total intensity were masked out of 
the Stokes $Q$ and $U$ images leaving pixels that are free of detectable polarized emission. 
For determining the noise statistics, areas near missing fields 
were not considered. The solid curves in Figure \ref{noisehisto-fig} represent 
Gaussians with standard deviation equal to the $rms$ of empty areas in the 
images avoiding sources. The non-Gaussian wings in $Q$ and $U$ emerging above 
the $2\sigma_{QU}$ level are related to the striping visible in 
Figure \ref{simulation-fig}. In both cases a Gaussian does not adequately 
represent the wings of the noise distribution. A better solution is the 
sum of a Gaussian and an exponential, 
\begin{equation}
\label{Gaussian-Exponential}
F(x) = A e^{-x^2 / 2 \sigma^{2}} + B e^{- C |x|} , 
\end{equation} 
with $A$, $B$ and $C$ determined from fitting. The parameters of the fit for the 
Stokes $Q$ simulated images are: $A=0.95341$, $B=0.00659$, $C=1.08135$ and 
$\sigma=0.44327$. 

To investigate the impact on false detection rate in polarized intensity, 
Monte-Carlo simulations of polarized intensity using the noise distribution 
from Equation \ref{Gaussian-Exponential} and $p_{0} = 0$ were done. 
First a $Q$ value was drawn. In principle $Q$ values were drawn independently 
from $U$ values. However, if the $|Q|$ was larger than a $2\sigma_{QU}$ threshold, $U$ 
values were drawn until the $|U|$ was also larger than the $2\sigma_{QU}$. 
This procedure acknowledges that in real data residual sidelobes in $Q$ 
probably also exist in $U$. 

\begin{figure}[h]
\begin{center}
\includegraphics[scale=0.40, angle=0]{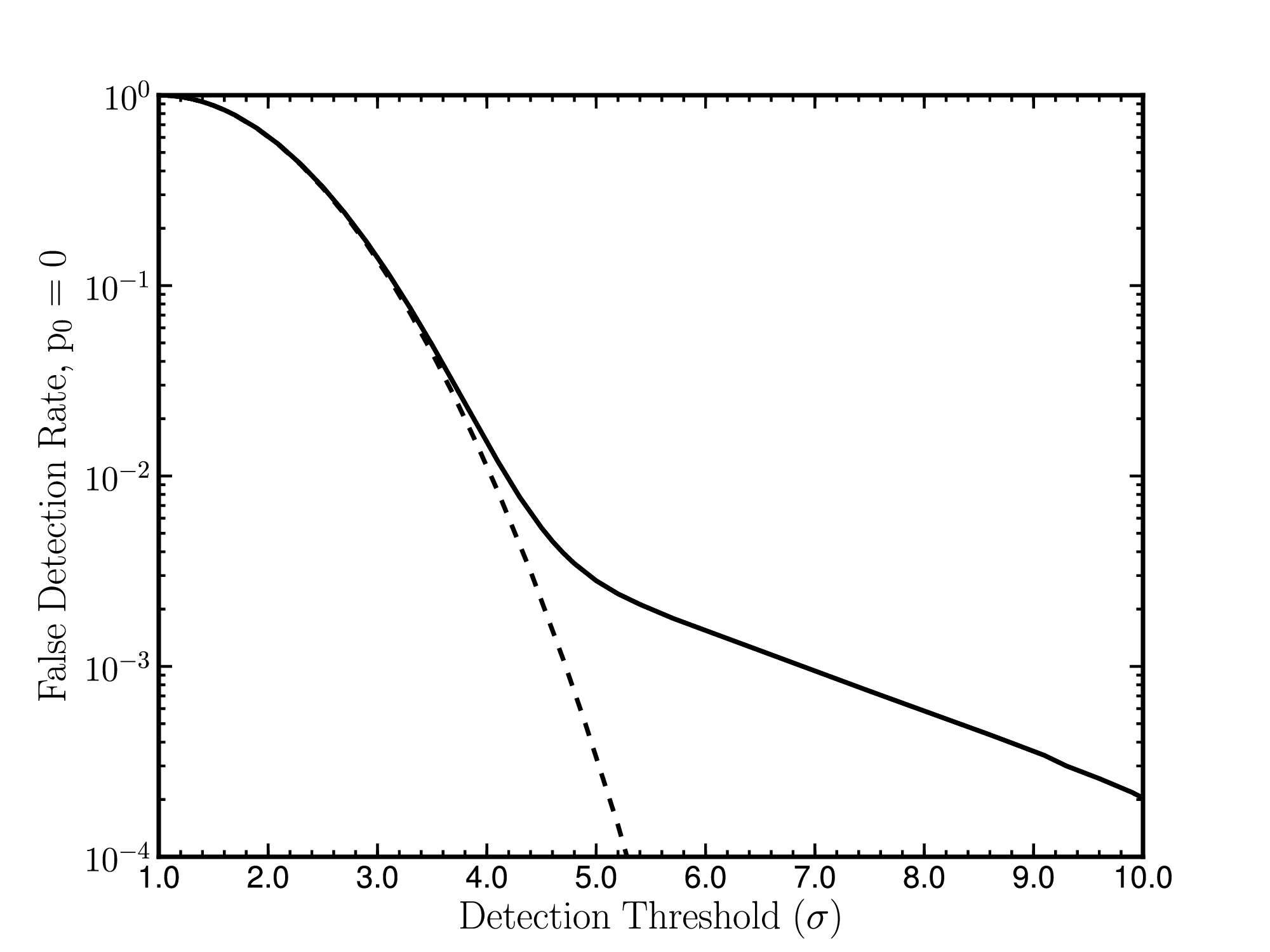}
\caption{False detection rates determined from Monte-Carlo simulations of the noise 
distribution for $p_{0} = 0$. The dashed curve represents a Gaussian fit 
to the noise and the solid curve uses the sum of the 
Gaussian and exponential noise function.}\label{pfalsedetect-fig}
\end{center}
\end{figure}

Figure \ref{pfalsedetect-fig} shows the false detection rate in polarized intensity 
for Gaussian and non-Gaussian noise in Stokes $Q$ and $U$ and Table \ref{FDRtable} lists 
false detection rates for a range of detection thresholds $p_{lim}$ in polarized intensity. 
The details of Figure \ref{pfalsedetect-fig} depend on the dynamic range of the $Q$ and $U$ 
images and will be different for every survey. The non-Gaussian wings in $Q$ and $U$ 
increase the false detection rate by orders of magnitude. In the simulations an $8\sigma$ 
detection threshold yields the same false detection rate as a $4.9\sigma$ 
detection threshold for Ricean statistics. 

Polarized source finding should apply a detection threshold that is derived from 
the actual noise distribution in $Q$ and $U$ of a dynamic range limited survey. 
We found no significant effect on the bias correction from the non-Gaussian wings. 

\begin{table}[h]
\begin{center}
\caption{False detection rates of polarized intensity for Gaussian (i.e. Ricean) and non-Gaussian (i.e. non-Ricean) noise in Stokes $Q$ and $U$. }\label{FDRtable} 
\begin{tabular}{lcc}
\hline $p_{lim}/\sigma_{QU}$ & Ricean & Non-Ricean \\
\hline 
3.0 & $1.36\times 10^{-1}$ & $1.41\times 10^{-1}$ \\
3.5 & $4.43\times 10^{-2}$ & $4.88\times 10^{-2}$ \\
4.0 & $1.13\times 10^{-2}$ & $1.51\times 10^{-2}$ \\
4.5 & $2.23\times 10^{-3}$ & $5.34\times 10^{-3}$ \\
5.0 & $3.35\times 10^{-4}$ & $2.82\times 10^{-3}$ \\
5.5 & $3.85\times 10^{-5}$ & $2.00\times 10^{-3}$ \\
6.0 & $2.90\times 10^{-6}$ & $1.54\times 10^{-3}$ \\
6.5 & $3.00\times 10^{-7}$ & $1.20\times 10^{-3}$ \\
7.0 & - & $9.43\times 10^{-4}$ \\
7.5 & - & $7.42\times 10^{-4}$\\
8.0 & - & $5.84\times 10^{-4}$ \\
8.5 & - & $4.60\times 10^{-4}$ \\
9.0 & - & $3.58\times 10^{-4}$ \\
9.5 & - & $2.73\times 10^{-4}$ \\
10.0 & - & $2.03\times 10^{-4}$ \\
\hline
\end{tabular}
\medskip\\
%%$^a$Table footnotes go here.\\
\end{center}
\end{table}

\section{Summary \& Conclusions}

The uncertainty in the Faraday depth of the source introduces a stronger bias 
in polarized intensity than just the well known polarization bias.
At low signal to noise the false detection rate is greatly enhanced, 
while at higher signal to noise ($p > 4\sigma_{QU}$) an effective 
estimator for the true polarized intensity is 
$\hat{p}_{0} = \sqrt{p^{2} - 2.3 \sigma_{QU}^{2}}$.

RM synthesis on the pixel nearest to the fitted position of total intensity 
introduces a systematic error that underestimates the polarized intensity by up to 
$15\%$. Source fitting in polarized intensity provides a more accurate result, 
even for the unresolved sources considered in this paper. 

Non-Gaussian wings of the noise distribution in Stokes $Q$ and $U$ significantly 
increase the rate of false detection in polarized intensity by orders of magnitude. 
False detection rates at $8\sigma_{QU}$ similar to Ricean false detection 
rates at $4.9\sigma_{QU}$.

%\section*{Acknowledgments} 

%\end{multicols}


\begin{thebibliography}{}

\bibitem[Beck \& Gaensler (2004)]{beck2004} Beck, R., \& Gaensler, B.~M., 2004, New Astronomy Reviews, 48, 1289

\bibitem[Brentjens \& De Bruyn(2005)]{brentjens2005} Brentjens, M.~A., \& De Bruyn, A.~G. 2005, A\&A 441, 1217

\bibitem[Burn(1966)]{burn1966} Burn, B.~J. 1966, MNRAS, 133, 67

\bibitem[Condon et~al. (1998)]{condon1998} Condon, J.~J., Cotton, W.~D., Greisen, E.~W., Yin, Q.~F., Perley, R.~A., Taylor, G.~B., \& Broderick, J.~J. 1998, AJ, 115, 1693

\bibitem[Gaensler et~al. (2010)]{gaensler2010} Gaensler, B.~M., Landecker, T.~L., Taylor, A.~R., POSSUM Collaboration, 2010, BAAS, 41, 515

\bibitem[Grant et~al. (2010)]{grant2010} Grant, J.~K., Taylor, A.~R., Stil, J.~M., Landecker, T.~L., Kothes, R., Ransom, R.~R., Scott, D., 2010, ApJ, 714, 1689

\bibitem[Heald (2009)]{heald2009} Heald, G., 2009, IAUS, 259, 591

\bibitem[Rice (1945)]{rice1945} Rice, S.~O. 1945, Bell Syst. Tech. J., 24, 46

\bibitem[Simmons \& Stewart(1985)]{simmons1985} Simmons, J.~F.~L., \& Stewart, B.~G. 1985, A\&A, 142, 100

\bibitem[Stil \& Taylor (2007)]{stil2007} Stil, J~M., Taylor, A.~R., 2007, ApJ, 663, L21

\bibitem[Subrahmanyan et~al. (2010)]{Subrahmanyan2010} Subrahmanyan, R., Ekers, R.~D., Saripalli, L., Sadler, E.~M., 2010, MNRAS, 402, 2792

\bibitem[Taylor et~al. (2007)]{taylor2007} Taylor, A.~R., Stil, J.~M., Grant, J.~K., Landecker, T.~L., Kothes, R., Reid, R.~I., Gray, A.~D., Scott, D., Martin, P.~G., Boothroyd, A.~I., Joncas, G., Lockman, F.~J., English, J., Sajina, A., Bond, J.~R. 2007, ApJ, 666, 201

\bibitem[Taylor \& Salter (2010)]{taylor2010} Taylor, A.~R., Salter, C.~J., 2010, Proceedings of the Conference "The Dynamic ISM: A Celebration of the Canadian Galactic Plane Survey", ASP Conference Series

\bibitem[Vaillancourt(2006)]{vaillancourt2006} Vaillancourt, J.~E. 2006, PASP, 118, 1340

\end{thebibliography}
\end{document}